\documentclass[a4paper,11pt]{article}
\pdfoutput=1 

\usepackage{jinstpub} 
\usepackage[active,new,old]{correct}

\title{\boldmath Channeling experiments at planar diamond and silicon single crystals with electrons from the Mainz Microtron MAMI}

\author[a,1]{H. Backe, \note{Corresponding author.}}
\author[a]{W. Lauth,}
\author[b]{and Thu Nhi TRAN THI}

\affiliation[a]{Institute for Nuclear Physics of Johannes Gutenberg-University, Johann-Joachim-Becher-Weg 45 \\D-55128 Mainz, Germany}
\affiliation[b]{X-ray Optics Group, ESRF, CS 40220\\F-38043 Grenoble Cedex 9, France}

\emailAdd{backe@uni-mainz.de}

\abstract{Line structures were observed for (110) planar channeling of electrons in a diamond single crystal even at a beam energy of 180 MeV. This observation motivated us to initiate dechanneling length measurements as function of the beam energy since the occupation of quantum states in the channeling potential is expected to enhance the dechanneling length. High energy loss signals, generated as a result of emission of a bremsstrahlung photon with about half the beam energy at channeling of 450 and 855 MeV electrons, were measured as function of the crystal thickness. The analysis required additional assumptions which were extracted from the numerical solution of the Fokker-Planck equation. Preliminary results for diamond are presented. In addition, we reanalyzed dechanneling length measurements at silicon single crystals performed previously at the Mainz Microtron MAMI at beam energies between 195 and 855 MeV from which we conclude that the quality of our experimental data set is not sufficient to derive definite conclusions on the dechanneling length. Our experimental results are below the predictions of the Fokker-Planck equation and somewhat above the results of simulation calculations of A. V. Korol and A. V. Solov'yov et al. on the basis of the MBN Explorer simulation package. We somehow conservatively conclude that the prediction of the asymptotic dechanneling length on the basis of the Fokker-Planck equation represents an upper limit.}

\keywords{Channeling phenomena, diamond, silicon single crystals, Fokker-Planck equation}

\proceeding{XII$^{\text{th}}$ International Symposium, RREPS-17\\
  September 18-22, 2017\\
  Hamburg, Germany}

\begin{document}
\maketitle
\flushbottom

\section{Introduction} \label{sec:intro}
An intriguing field of research is intense radiation production with photon energies of 100 keV or more employing channeling of relativistic positrons or electrons with energies in the order of a few hundred MeV or higher. A very important prerequisite is the knowledge of
the dechanneling length, i.e., the length before a charged
particle is kicked out by a collision with an atom from the channel. Of particular interest are electrons since high quality electron
beams can, in comparison with positron beams, much easier be supplied.

Currently simulation calculations are widely used to get information on dechanneling lengths also for bent crystals, see, e.g., the recent articles of A. V. Korol, A. V. Solov'yov et al. \cite{KorB16, KorB17}. Experimentally there are two possibilities to measure a dechanneling length.
The first one is based on a variation of the thickness of straight crystals, the second one on the observation of dechanneled electrons in a bent crystal \cite{MazB14, WisU16}. This paper deals with the former possibility which we employed in previous experiments at MAMI to determine the dechanneling length for electrons in (110) silicon single crystals \cite{LauB08, LauB10}. Such kind of measurements were analyzed under various assumptions \cite{BacL08, BacK12, BacL15} which resulted in different dechanneling lengths.

In this paper we describe the continuation of such measurements on diamond single crystals at (110) planar channeling. Our main interest was whether quantum effects could enhance the dechanneling length at electron beam energies below about 200 MeV. In the first part of this paper we describe a measurement of the photon spectrum taken at (110) channeling at 180 MeV. This section is followed by a measurement of the dechanneling length at 450 and 855 MeV. The novel analysis method developed for this experiment has been employed also for a reanalysis of our previous measurements for (110) channeling at silicon single crystals.

\section{Measurement of the photon spectrum for (110) planar channeling of 180 MeV electrons at diamond} \label{diamondSpectra}
The results of previous dechanneling length measurements for (110) silicon single crystals suggested that the dechanneling length enhances at beam energies below 200 MeV when compared with various theoretical predictions \cite{BacL15}. These findings were interpreted to originate from the occupation of deeply bound quantum states which require a rather large scattering angle in a single scattering event to achieve the energy transfer for dechanneling. Such processes may happen only in the tails of the scattering distribution function, and are, therefore, expected to be suppressed.

\begin{figure}[tb]
\centering
    \includegraphics[angle=0,scale=0.20,clip] {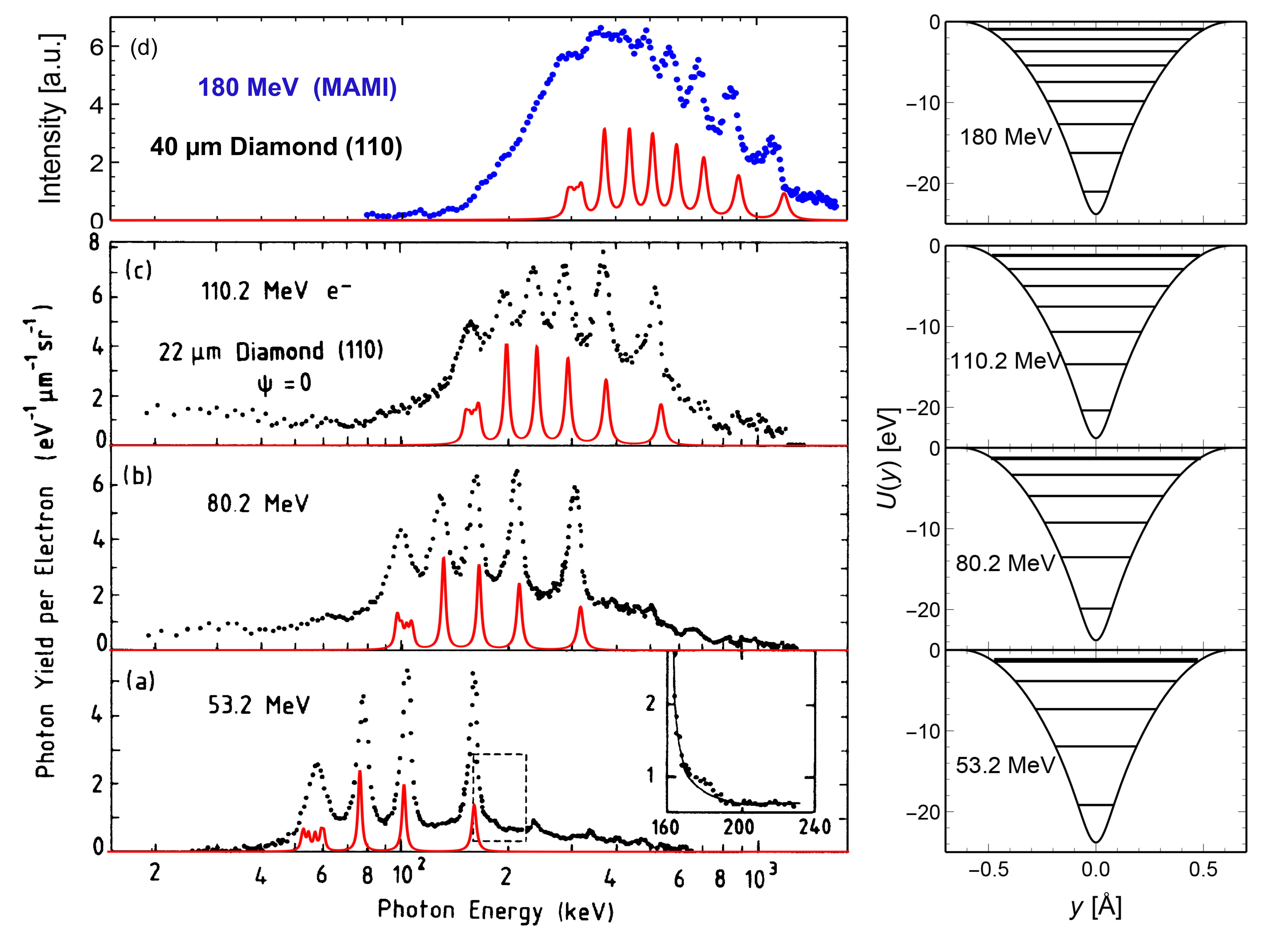}
\caption[] {Photon spectra at (110) planar channeling of electrons in diamond. Spectra (a)-(c) reproduced from M. Gouanere et al. \cite{GouS88}, (d) deconvoluted photon spectrum taken at MAMI with a Ge(i) detector at channeling of 180 MeV electrons in a 40 $\mu$m thick diamond single crystal (preliminary). The bremsstrahlung contribution has been subtracted. For experimental setup and procedure see \cite{BacK12}. Right hand side: Potentials and level structures as calculated with the computer code of B. Azadegan \cite{Aza13}. The line structures as calculated with the same code for a 40 $\mu$m thick crystal at an incident angle of 300 $\mu$rad (lower red curves) are systematically too high in energy, indicating that the potential in the approximation of  P.A. Doyle and P.S. Turner \cite{DoyT67} must obviously be modified. The fact that channeling experiments are sensitive to the potential shape is well known, see e.g. \cite{PanK87, PatS87}. } \label{DiamondLineSpectrum180MeV}
\end{figure}

The existence of quantum states is well known from the observation of line structures at (110) channeling of electrons in diamond single crystals at beam energies between 53.2 to 110.2 MeV \cite{GouS88}. We supplemented the investigation with a measurement at a beam energy of 180 MeV. The deconvoluted spectrum is depicted in figure \ref{DiamondLineSpectrum180MeV} (d).
It shows that even at such a high beam energy line structures are present. This observation motivated us to initiate dechanneling length measurements as function of the beam energy also for diamond.

\section{Dechanneling length measurements for diamond single crystals} \label{dechannelingStraightMAMI}
\begin{figure}[tbh]
    \includegraphics[angle=0,scale=0.45,clip]{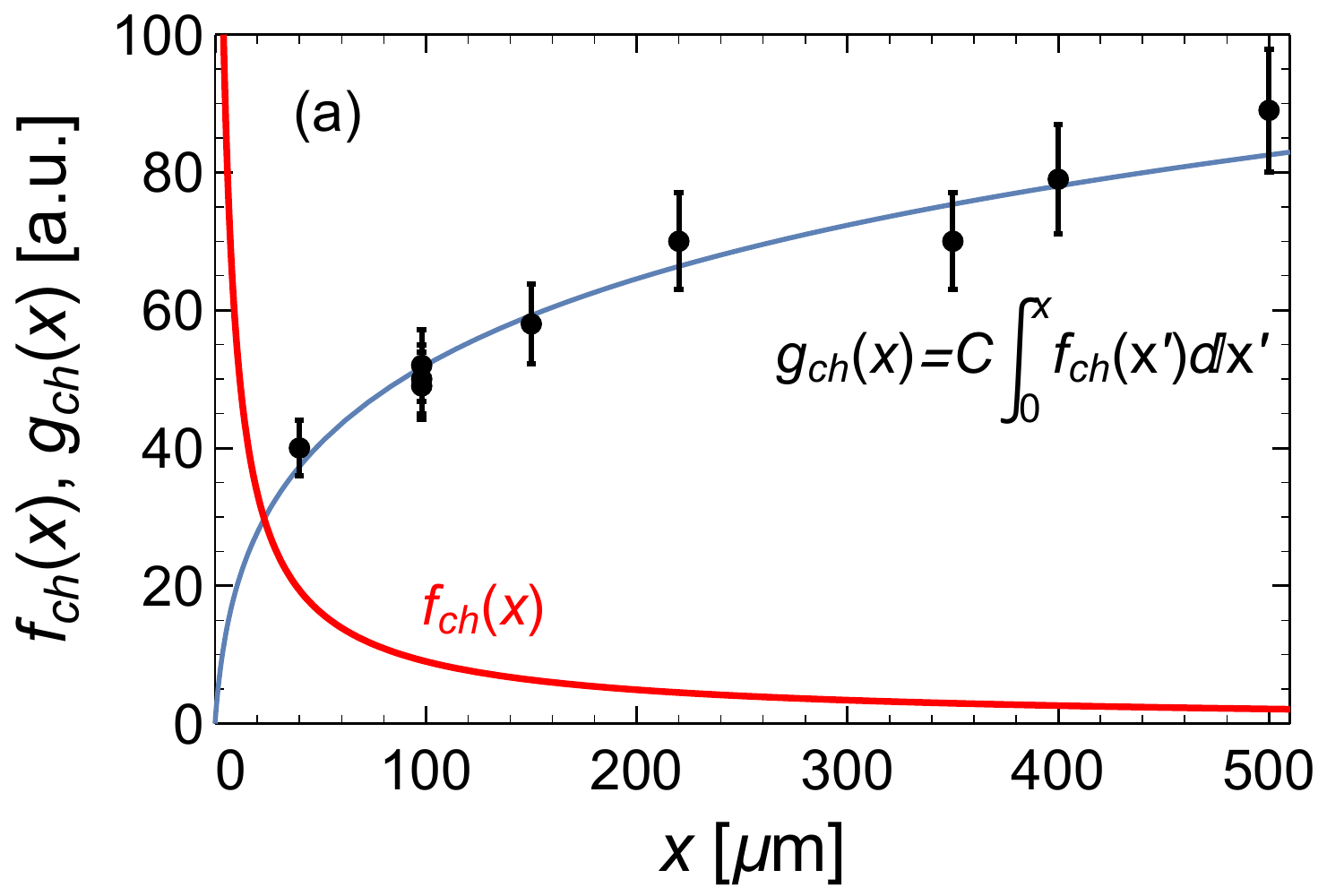}
    \hspace*{0.4cm}
    \includegraphics[angle=0,scale=0.45,clip]{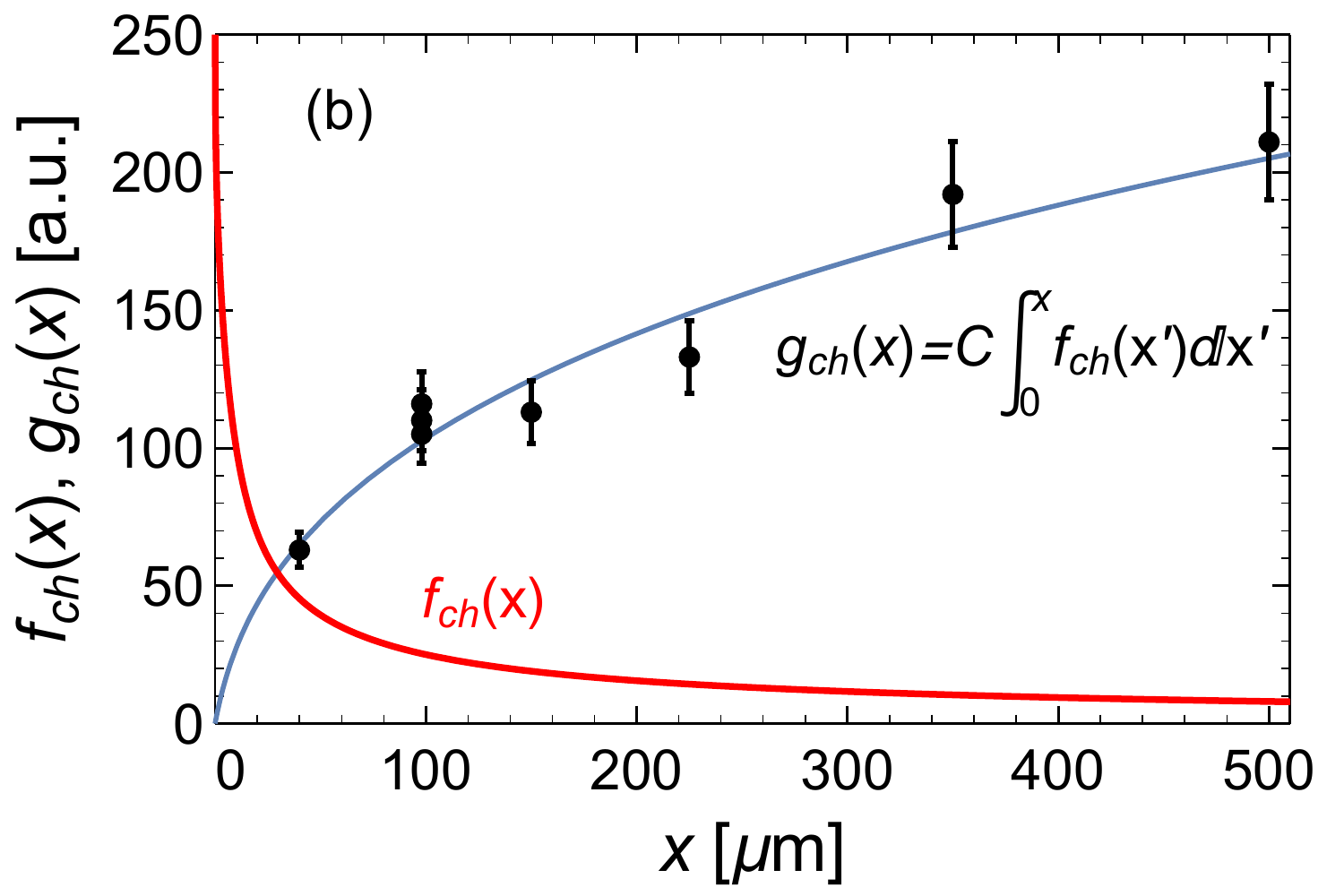}
\caption[]{High energy loss signals as function of the crystal thickness for planar channeling of 450 MeV electrons (a) and 855 MeV electrons (b) in (110) diamond single crystals (preliminary). The error bars take into account the quality of the crystals in terms of the dislocation density. The red curve denoted by $f_{ch}(x)$ is the solution of equation (\ref{channelOccupODE}) for the best fit with the function ${g_{ch}}(x) = C\int_0^x {{f_{ch}}(x')\;dx'}$ (blue curve) to the experimental data. Fit parameters are $A$ and $C$.} \label{dechannelingDiamond}
\end{figure}

In order to investigate further whether the observation of line structures even at 180 MeV supports the picture of an enhanced dechanneling length, or not, we performed measurements for (110) planar channeling at diamond with the method described in \cite{BacK12A}. The high energy loss signals generated at emission of a bremsstrahlung photon at channeling with an energy $\hbar\omega \approx E/2$, $E$ is the beam energy, are depicted for various crystal thicknesses between 40 and 500 $\mu$m by the error bars in figure \ref{dechannelingDiamond}. The signal is assumed to be proportional to the integral $\int_0^x {{f_{ch}}(x')\;dx'}$ with $f_{ch}(x')$ the channel occupation probability. The question is now how to extract from the channel occupation probability as function of the crystal thickness a dechanneling length. Taking into account also rechanneling, one might try an ansatz with the aid of the ordinary differential equation
\begin{eqnarray}
{f_{ch}}'(x) + {f_{ch}}(x)~A \cdot \big( {{\lambda _{de}}(x) - {\lambda _{re}}(x)} \big) = 0.
\label{channelOccupODE}
\end{eqnarray}
The problem with this ansatz is that the occupation probability depends, beside the dechanneling rate $\lambda_{de}(x)$ which is the quantity of interest, on equal footing also on the rechanneling rate $\lambda_{re}(x)$. Obviously, the dechanneling rate cannot be extracted without additional assumptions. Because of lack of better theoretical data we utilized the solutions of the Fokker-Planck equation. How this has been performed will be described in the next section.

\section{Fokker-Planck equation for a plane crystal}
\label{FokkerPlanck}
In the analysis of our experiments we take advantage of numerical solutions of the Fokker-Planck equation, see e.g. \cite{BaiK98, BelT81, KumK89}. Well adapted for our purpose are descriptions in \cite{BacL08, BacK12, BacL15}. A coordinate system has been chosen with the $x$ axis pointing into the initial beam direction, and the $y$ axis perpendicular to the channeling planes. The Fokker-Planck equation with $F(x/L_{de}, E_{\perp}/U_0) = F(\xi,\varepsilon_{\perp})$ the probability density and $J(x/L_{de}, E_{\perp}/U_0) = J(\xi,\varepsilon_{\perp})$ the probability current written with the normalized variables $\xi = x/L_{de}$ and $\varepsilon_{\perp}= E_{\perp}/U_0$ reads with the diffusion coefficient $D_{diff}^{(2)}(\varepsilon_{\perp})$ and the time period $T(\varepsilon_{\perp})$
\begin{eqnarray}
\frac{\partial F(\xi,\varepsilon_{\perp})}{\partial \xi}=-\frac{\partial J(\xi,\varepsilon_{\perp})}{\partial\varepsilon_{\perp}}= \frac{\partial}{\partial \varepsilon_{\perp}} \Big[D_{diff}^{(2)}(\varepsilon_{\perp})T(\varepsilon_{\perp})\frac{\partial}{\partial \varepsilon_{\perp}}\frac{F(\xi,\varepsilon_{\perp})}{T(\varepsilon_{\perp})}\Big], \label{FokkerPlanck1}
\end{eqnarray}
\begin{eqnarray}\label{KitagavaOhtsuki}
D_{diff}^{(2)}(\varepsilon_{\perp}) =
\frac{2}{T(\varepsilon_{\perp})\cdot c} \int\limits_{y_{min}}^{y_{max}}\frac{d_{p}}{\sqrt{2\pi}u_{1}}
\exp(-\eta^{2}/2u_{1}^{2})\sqrt{2\big(\varepsilon_{\perp}-u(\eta)\big)\cdot\gamma
m_e c^2/U_0}~d\eta,
\end{eqnarray}
\begin{eqnarray}\label{Period}
T(\varepsilon_{\perp})\cdot c = 2\int\limits_{y_{min}}^{y_{max}} \sqrt{\frac{\gamma m_e c^2/U_0}{2\big(\varepsilon_{\perp}-u(y)\big)}}~~dy.
\end{eqnarray}
The integration limits $y_{min}$ and $y_{max}$ are the roots of the equation $E_{\perp}/U_0=\varepsilon_{\perp}=u(y)=U(y)/U_0$. The two normalizing variables are the depth $U_0$ of the potential $U(y)$, and the scaling length
\begin{eqnarray} \label{LdeBaier}
L_{de} = 2\frac{{{U_0}\;pv\;{X_0}}}{{E_s^2}}.
\end{eqnarray}
The quantity $X_{0}$= 0.1213 m is the radiation length \cite{Pat17}, $p$ the momentum of the electron, $E = \gamma m_{e} c^{2}$ its total energy, c the speed of light, $\gamma=1/\sqrt{1-(v/c)^2}$, $v$ the velocity of the electron, and $u_{1}=\sqrt{B/8\pi^2}$ = 0.0423 \AA~ the standard deviation of the one dimensional thermal vibration amplitude at 293 K with $B=0.141$~\AA$^2$ according to \cite[equation (15)]{SeaS91}. The quantity $d_p = a/2 \sqrt{2}$ = 1.261 \AA~ in the nominator of the integral in equation (\ref{KitagavaOhtsuki}) is the (110) interplanar distance, with $a$ = 3.567 \AA~ the lattice constant at 300 K, taken from \cite{Ioffe, Matprop}. All numbers are given for diamond.

\begin{figure}[tbh]
\centering
    \includegraphics[angle=0,scale=0.45,clip]{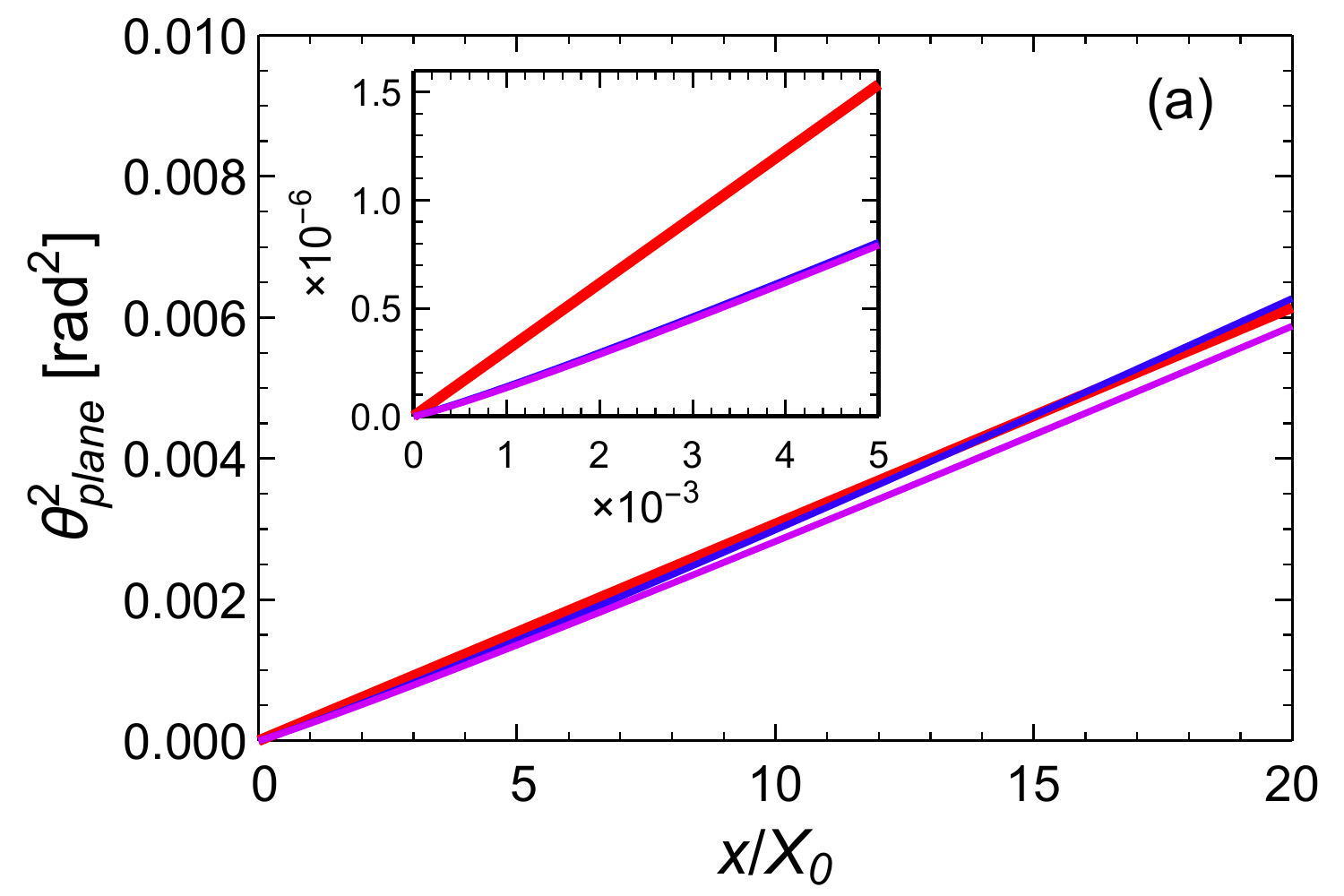}
    \includegraphics[angle=0,scale=0.45,clip]{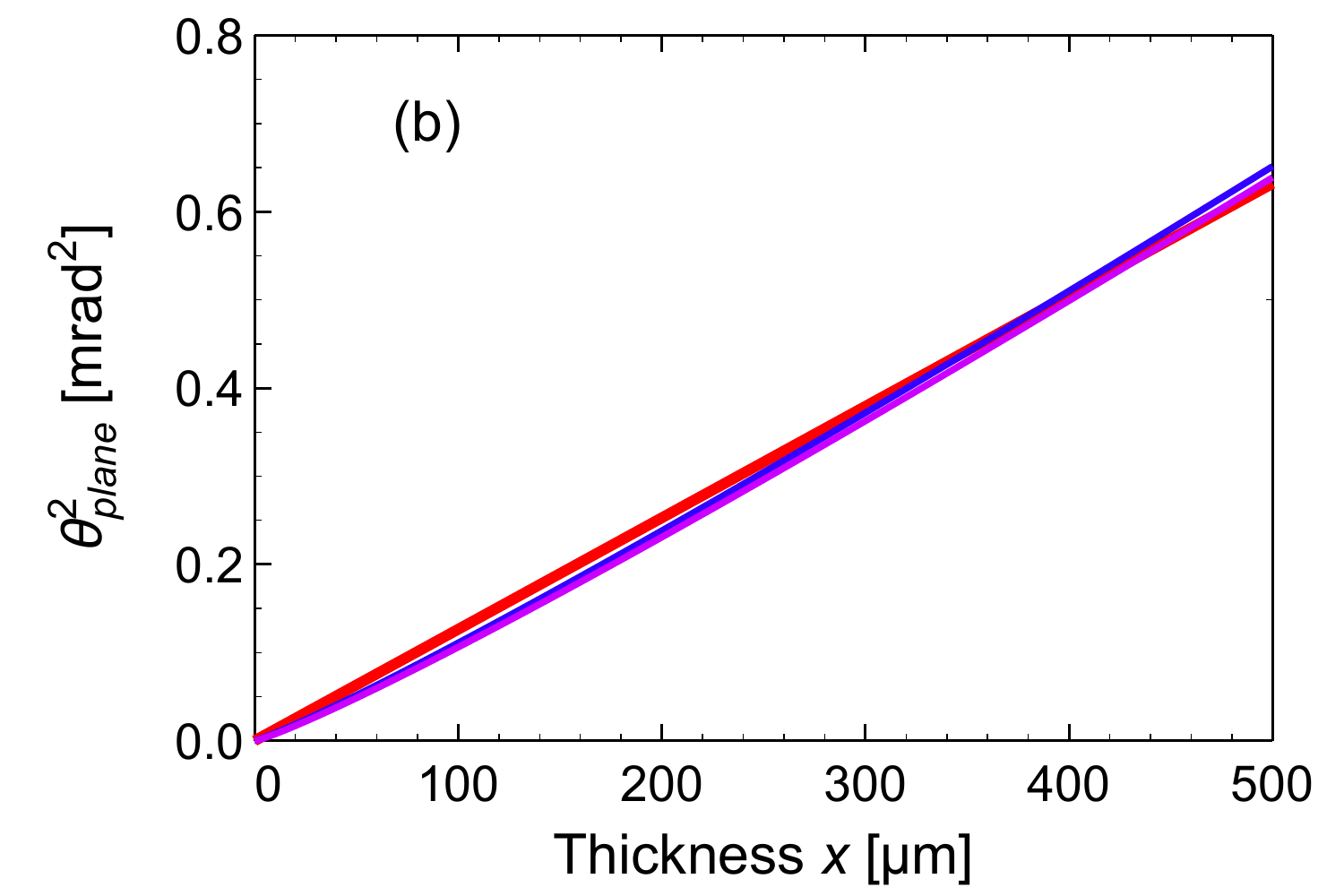}
\caption[]{(a) Variances of Gaussian scattering distributions as function of thickness $x$ in units of the radiation length $X_0= 121300~\mu$m for diamond. Red curve Rossi-Greisen approximation \cite[\S 22 "Multiple Scattering...."]{RosG41} $\theta _{plane}^2 = {\left(E_{s}/pv \right)^2}\cdot (x/X_0)$ with $E_{s}^2 = (2\pi/e^2)\cdot m^2 \Rightarrow (2\pi/\alpha)\cdot\left( {m_e}{c^2} \right)^2 = {\left( {15.0\;{\rm{MeV}}} \right)^2}$ as specified by Baier et al. \cite[chapter 10.1]{BaiK98}, blue curve variance of particle data group with $\theta _{plane}^2=\theta _{0}^2=(13.6~\mbox{MeV}/pv)^2\cdot (x/X_0)\big[1+0.038 \ln(x/X_0)\big]^2$, and violet curve variance of G.R. Lynch and O.I. Dahl \cite{LynD91} with $F = 0.98$ the fraction of tracks in the sample. (b) With $E_s = 10.6$ MeV the red curve approaches for $x < 500~\mu$m or $x/X_0<4.1\cdot 10^{-3}$ the blue and violet ones.} \label{theta2rms}
\end{figure}

Equation (\ref{LdeBaier}) is exactly the expression which Baier et al. quote for the dechanneling length \cite[eqn. (10.1)]{BaiK98}, however, here with a modified $E_s$ = 10.6 MeV valid for small thicknesses $x \ll X_0$. For the reasoning see figure \ref{theta2rms}.

\begin{figure}[tbh]
\centering
    \includegraphics[angle=0,scale=0.55,clip]{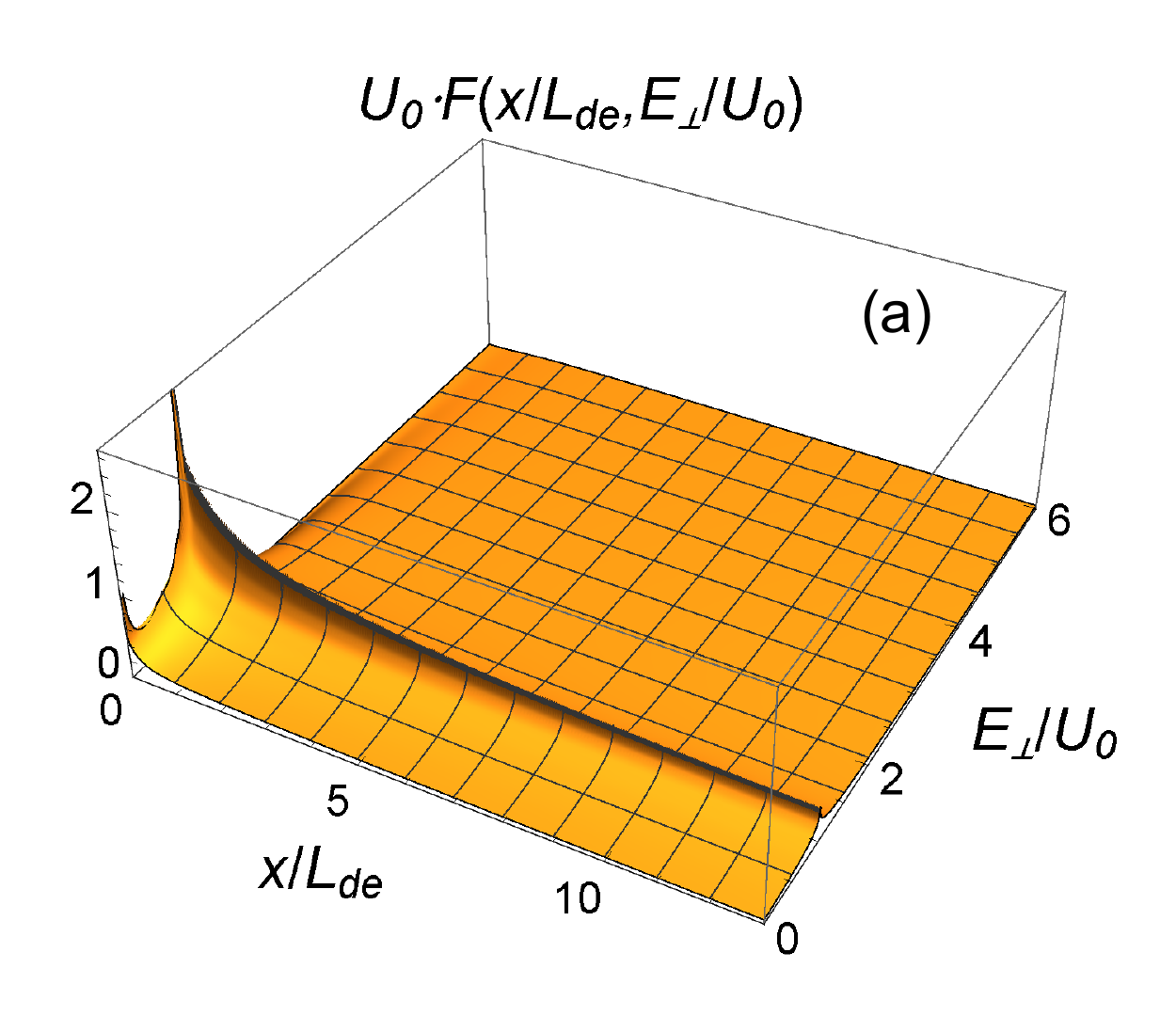}
    \includegraphics[angle=0,scale=0.5,clip]{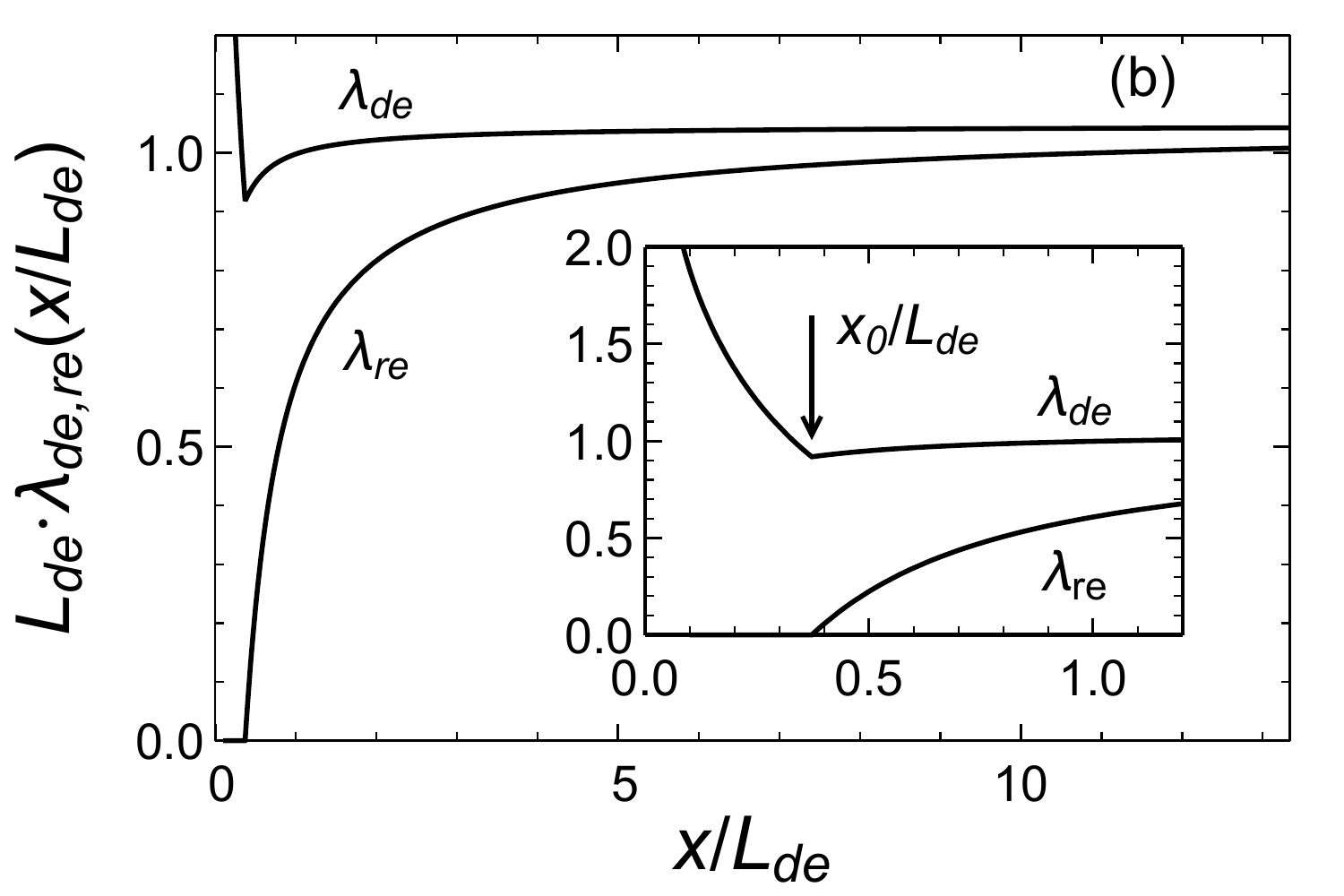}
\caption[]{Numerical solution of the Fokker-Planck equation depicted in normalized variables. The probability density (a) and the transition rates (b) were calculated with equations (\ref{lambdaDeFP}) and (\ref{lambdaReFP}). For the electron beam a Gaussian angular beam profile was assumed with a standard deviation of $\sigma_y ' = 30~\mu$rad, entering the crystal parallel to the (110) planes. Here are $x_0/L_{de} = 0.372$, $L_{de}\cdot\lambda_{de}(x/L_{de} \rightarrow 13.33)= 1.0424$, and $L_{de}\cdot\lambda_{re}(x/L_{de} \rightarrow 13.33)= 1.0045$.} \label{FokkerPlanckSolutions}
\end{figure}

From a solution of the Fokker-Planck equation the channel occupation probability can be calculated with the integral
\begin{eqnarray}
f_{ch}(\xi) = \int_{0}^{1} F(\xi,\varepsilon_{\perp})~d\varepsilon_{\perp}.
\label{channelOccupation}
\end{eqnarray}
It depends on the dechanneling and rechanneling currents $J_\uparrow(\xi,{\varepsilon_{\perp}} = 1)$ and $J_\downarrow (\xi,{\varepsilon_{\perp}} = 1)$ across the continuum border from which transition rates for de- and rechanneling rates can be calculated as
\begin{eqnarray}\label{lambdaDeFP}
{\lambda _{de}}(x/L_{de})={\lambda _{de}}(\xi) = \frac{{{J_ \uparrow }(\xi,{\varepsilon_{\perp}} = 1) }}{{{f_{ch}}(\xi)}}=\frac{{{J_{drift}}(\xi,{\varepsilon_{\perp}} = 1) + \theta (x_0/L_{de}-\xi)\;{J_{diff}}(\xi,{\varepsilon_{\perp}} = 1) }}{{{f_{ch}}(\xi)}}
\end{eqnarray}
\begin{eqnarray}\label{lambdaReFP}
{\lambda _{re}}(x/L_{de})={\lambda _{re}}(\xi) = - \frac{{{J_ \downarrow }(\xi,{\varepsilon_{\perp}}=1 ) }}{{{f_{ch}}(\xi)}}= - \theta (x_0/L_{de}-\xi)\;\frac{{{J_{diff}}(\xi,{\varepsilon_{\perp}} = 1)}}{{{f_{ch}}(\xi)}}.
\label{lambdaRe}
\end{eqnarray}
By means of the relation $\partial [T(\varepsilon_{\perp})D_{diff}^{(2)}(\varepsilon_{\perp})]/\partial \varepsilon_{\perp}=T(\varepsilon_{\perp})D_{drift}^{(1)}(\varepsilon_{\perp})$, which also defines the drift coefficient $D_{drift}^{(1)}(\varepsilon_{\perp})$, the current $J(\xi,\varepsilon_{\perp})$ in equation (\ref{FokkerPlanck1}) has been subdivided into a diffusion $J_{diff}(\xi,\varepsilon_{\perp})$ and a drift current $J_{drift}(\xi,\varepsilon_{\perp})$:
\begin{eqnarray}\label{FokkerPlanckCurrent}
J(\xi,\varepsilon_{\perp}) = -\frac{\partial}{\partial \varepsilon_{\perp}} \Big[D_{diff}^{(2)}(\varepsilon_{\perp})F(\xi,\varepsilon_{\perp})\Big]+ D_{drift}^{(1)}(\varepsilon_{\perp}) F(\xi,\varepsilon_{\perp})= J_{diff}(\xi,\varepsilon_{\perp})+J_{drift}(\xi,\varepsilon_{\perp}).
\label{FokkerPlanck2}
\end{eqnarray}
The Heaviside $\theta$ function takes into account that for $x/L_{de}\leq x_0/L_{de}=0.372$ the diffusion current is outward directed, i.e., it contributes to the dechanneling rate with vanishing rechanneling, see figure \ref{FokkerPlanckSolutions}. 
It appears that for small thicknesses an equilibration phase exists which depends on the angular spread $\sigma_y '$ of the incoming electron beam and lasts for the example of $\sigma_y ' = 30~\mu$rad about one dechanneling length, $x/L_{de}\approx 1$. From this discussion we may conclude that the solution for small penetration depths $x/L_{de} \lessapprox 1$ is probably rather inaccurate.

Several other interesting features can be recognized from figure \ref{FokkerPlanckSolutions} (b). Most important is that the dechanneling rate scales in both coordinates with the scaling length $L_{de}$. Asymptotically the normalized de- and rechanneling rates approach numbers which are close to, but not exactly, unity, for numerical values see caption of figure \ref{FokkerPlanckSolutions}. We further mention that for large $x/L_{de}$ the absolute values of the transition rates approach each other with the diffusion rate always somewhat smaller. Qualitatively it can be concluded from equation (\ref{LdeBaier}) that for a large dechanneling length the beam energy $pv$, potential depth $U_0$, and the radiation length $X_0$ all should be large.

\section{Analysis procedure and results}
\label{Analysis}
\begin{figure}[tbh]
\centering
   \includegraphics[angle=0,scale=0.44,clip]{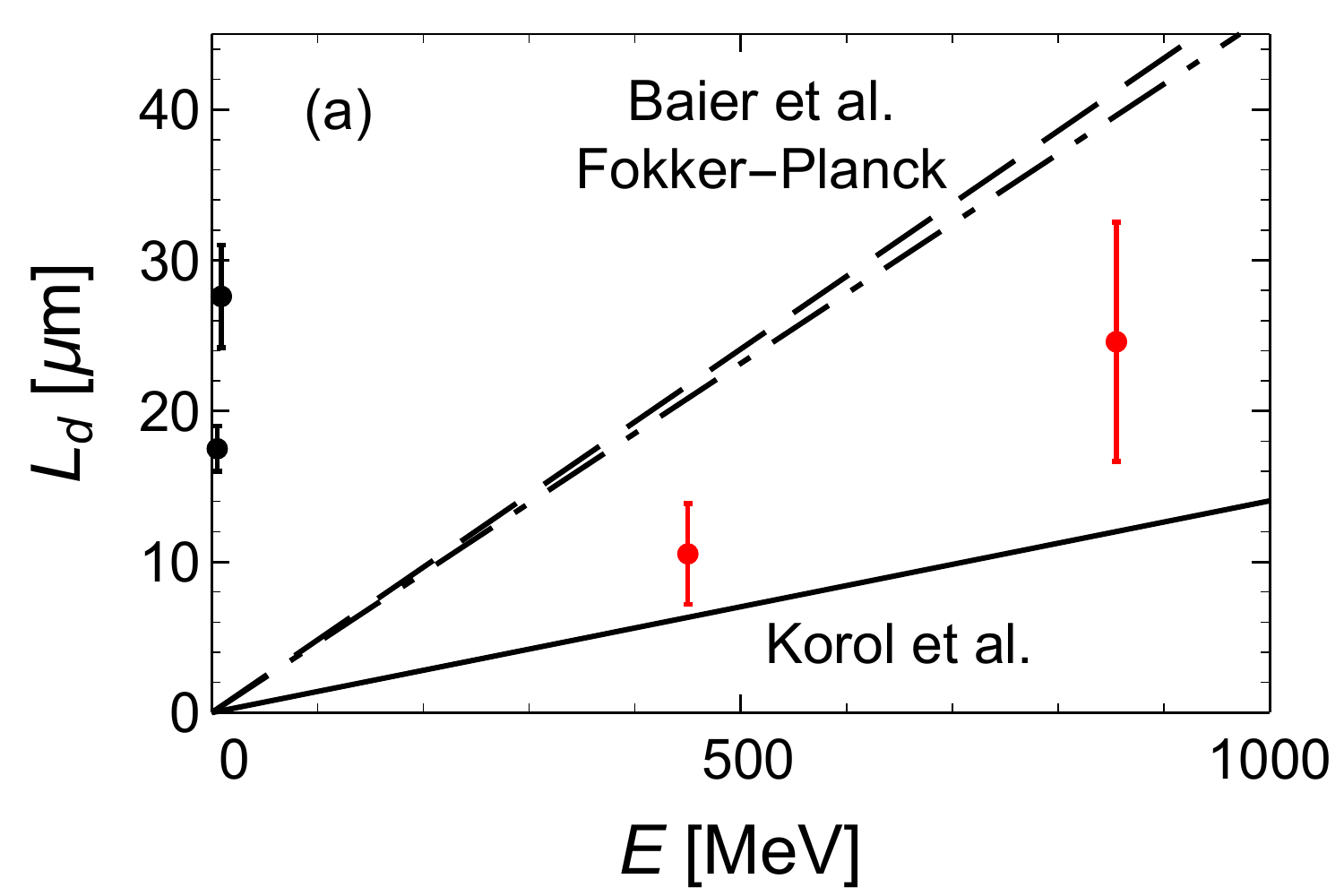}
    \hspace*{0.0cm}
    \includegraphics[angle=0,scale=0.44,clip]{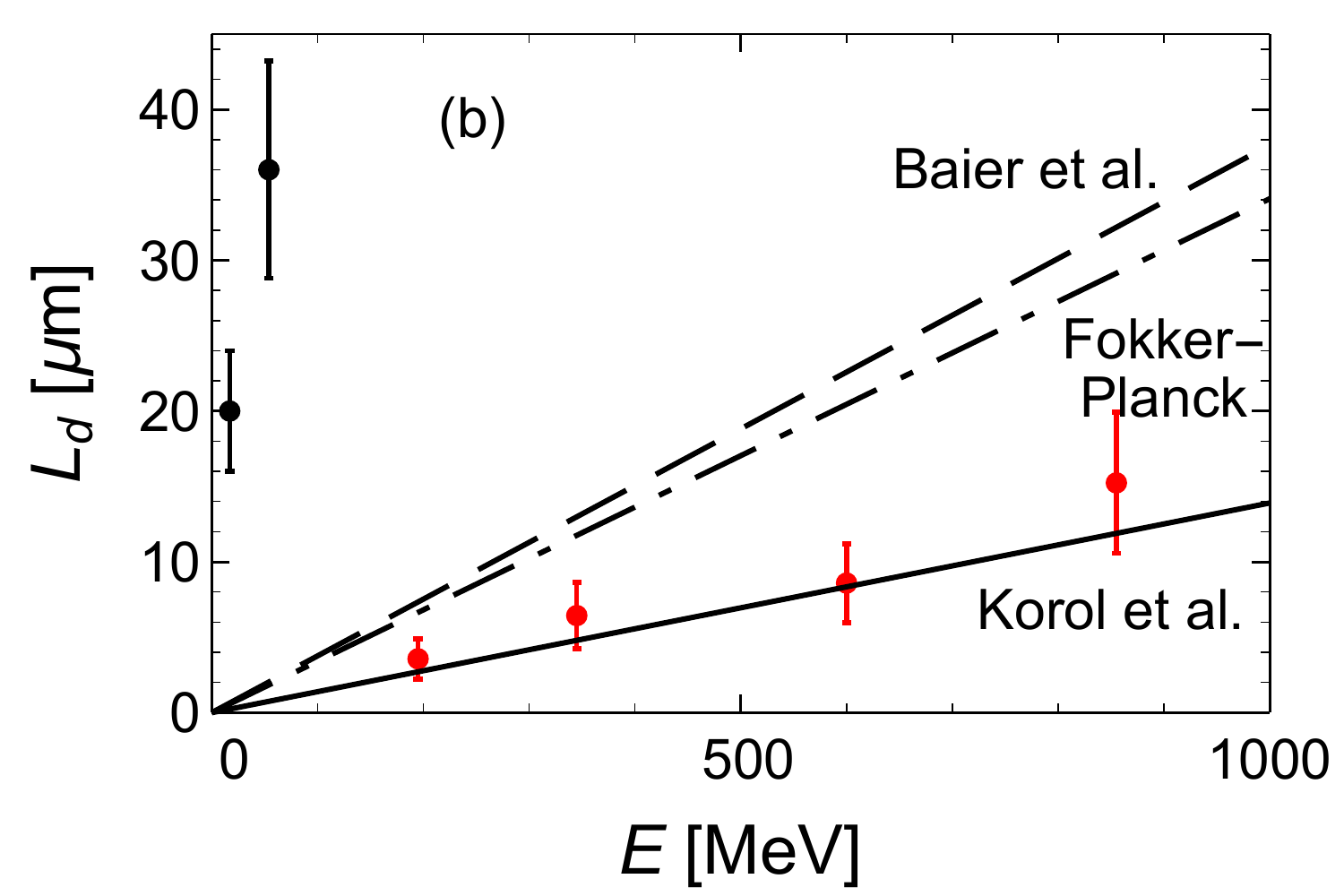}
\caption[]{(a) Asymptotic (110) dechanneling lengths as function of the beam energy for diamond. (Numerical values of depicted results are $L_d$ = 10.5(3.3) and 24.6(7.9) $\mu$m for 450 and 855 MeV, respectively.) A systematic error of 30 \% has been quadratically added to the fit errors. Data points at 5.2 and 9.0 MeV taken from \cite{NetG94}, curve denoted by Korol et al. derived from a calculation (12.01 $\pm$ 0.40)~$\mu$m at 855 MeV \cite[Table 1]{KorB17} by linear interpolation to the origin. (b) Same analysis for (110) planar channeling of electrons at silicon, data taken from \cite{BacL15}. (Numerical values of depicted results are $L_d$ = 3.6(1.3), 6.4(2.2), 8.6(2.6), and 15.2(4.7) $\mu$m for 195, 345, 600 and 855 MeV, respectively.) Also here a systematic error of 30 \% was quadratically added to the fit errors. Data points at low beam energies of 17 and 54 MeV taken from \cite{KepP89}. Curve denoted by Korol et al. derived from a calculation (11.89 $\pm$ 0.49)~$\mu$m at 855 MeV \cite[Table 1]{KorB16} by linear interpolation to the origin.} \label{dechannelingLengts}
\end{figure}

In the analysis we assumed that the difference ${{\lambda _{de}}(x) - {\lambda _{re}}(x)}$ can be taken from the Fokker-Planck equation. This assumption may well be allowed for the beam eneries of 450 and 855 MeV for which quantum state effects most likely are not of importance. The factor $A$ in equation (\ref{channelOccupODE}) has been treated as a fit parameter. The experimental dechanneling length is $L_{d}^{exp}=1/(A \lambda_{asymp})$ with $\lambda_{asymp} \simeq \lambda_{de} (x \rightarrow 500 \mu\mbox{m})$. In figure \ref{dechannelingDiamond} (a) and (b) the results of the best fits procedures are shown. Figure \ref{dechannelingLengts} (a)  depicts the dechanneling lengths for the two beam energies of 450 and 855 MeV. Because of technical reason during the course of the experiment we could not take data for the most interesting beam energy of 180 MeV. Therefore, we also reanalyzed with the same method older measurements for (110) channeling at silicon single crystals. The results are shown in figure \ref{dechannelingLengts} (b). The conjectured effect was not found.

It can be seen from figure \ref{dechannelingLengts} (a) and (b) that large deviations of the theoretical predictions for the dechanneling lengths exist  which, consequently, are also expected for the difference in the transition rates $\lambda _{de}(x) - \lambda _{re}(x)$. However, such a possibility is partly taken into account in the best fit procedure by the parameter $A$ in equation (\ref{channelOccupODE}). As mentioned, for thicknesses $x<L_{de}$ the accuracy of $\lambda _{de}(x) - \lambda _{re}(x)$ obtained from the Fokker-Planck equation might be questionable. To estimate the sensitivity, the rechanneling curve has been shifted by $\pm$ $x_0/L_{de}$ which resulted in a change of the asymptotic dechanneling length in the order of $\pm$~30 \% which has been taken into account in the error bars of figure \ref{dechannelingLengts}.

\section{Discussion and conclusion}\label{discussion}
Some remarks of caution on our previous measurements for (110)
channeling at silicon for beam energies between 195 and 855 MeV and the
analysis of these data may be appropriate. We would like to stress that the
signals observed for diamond shown in figure \ref{dechannelingDiamond} do not
saturate as function of the crystal thickness what was not observed for
silicon \cite{LauB08,LauB10}. Since we concluded from such a saturation
evidence for quantum states effects, the experiment on diamond casts serious
doubts on the assumptions made in our previous analysis \cite{BacL15}. Assuming that the classical picture must be favored, our current
model assumptions seem to be more realistic. However, we must concede that much more accurate experimental data are required to validate
on the basis of a $\chi^2$ analysis certain model assumptions, or exclude
others. Also the experiment on diamond, see figure \ref{dechannelingDiamond}, must be improved. In this context we mention that the various crystals used for the measurement may scatter in quality, meaning that, e.g., the dislocation density may vary from crystal to crystal, and in turn also the signals obtained for a dechanneling length measurement. Such a possibility has been partly included into the error bars in figure \ref{dechannelingDiamond}.

The most striking feature of the theoretical curves in figure  \ref{dechannelingLengts}, black lines, is the fact that the dechanneling lengths on the basis of the Fokker-Planck equations (\ref{FokkerPlanck1}) and the simulation calculations of Korol et al. \cite{KorB16, KorB17} disagree by a large factor. We can discuss this deviation on a safe ground only for the solution of the Fokker-Planck equation.

First of all one might question the prerequisite of the Fokker-Planck equation for which statistical equilibrium has been assumed meaning that the probability distribution in a channel is represented by
\begin{eqnarray}\label{ProbabilityDistribution}
\frac{dP}{dy}(y, \varepsilon_{\perp})=\frac{2}{T(\varepsilon_{\perp})\cdot c}\sqrt{\frac{\gamma m_e c^2/U_0}{{2\big(\varepsilon_{\perp}}-u(y)\big)}}.
\end{eqnarray}
For $\varepsilon_{\perp} = E_{\perp}/U_0 \leq 1$ this function has singularities which turn out to be harmless for the calculation of the diffusion coefficients or for the solution of the Fokker-Planck equation. In particular, the probability currents which enter the transition rates are smooth as function of $\varepsilon_{\perp}$. However, the singularities push the probability density towards the potential walls resulting in particular for deeply bound electrons in a reduction at the center where the scattering centers are located. The effect is rather large, e.g., for $\varepsilon_{\perp}=0.5$ the total probability reduces from 1 to 0.67 at integration between 90 \%  of $y_{min}$ and $y_{max}$. As a consequence, the calculated diffusion coefficients which enter into the Fokker-Planck equation may be too small, and in turn also the transition rates.

Further effects will be discussed by means of the scaling length $L_{de}$, equation (\ref{LdeBaier}). We recognize that the most sensitive parameter is $E_s$ since it enters quadratically. To explain a factor of even 2.5 we would have to assume an $E_s = 16.6$~MeV which can be ruled out by the same argumentation carried out above in figure \ref{theta2rms} (inset). There remains the potential depth $U_0$ to be discussed. The potential $U(y)$ has been calculated in the Doyle-Turner \cite{DoyT67} and in the Moli\`{e}re approximation, the latter with the parameter set of \cite[p.199 and 235]{BaiK98}. We calculated  for $U_0$ values of 22.41 eV (Moli\`{e}re) and 23.42 eV (Doyle-Turner) for (110) diamond, and 22.61 eV (Moli\`{e}re) and 21.09 eV (Doyle-Turner) for (110) silicon, i.e., the agreement is better than 7.2\%. Therefore, it is hard to conceive that the potentials are wrong by about a factor of two.

From figure \ref{FokkerPlanckSolutions} we see already qualitatively that a large fraction of electrons are weakly bound. Indeed 24 \% of all electrons have a binding energy of less than 10 \% of $U_0$. Those electrons may be excited into the continuum by electron-electron interactions. The latter is not included into $E_s$, but it has been mentioned in \cite{BacL15} that the effect on the dechanneling length would be in the 10 \% range. However, the electron density between the planes may well be different as it was calculated from the Moli\`{e}re potential. We also mention an additional idealization namely that in the Moli\`{e}re potential the microscopic atomic structure of a plane is averaged out. Any wavy structure may cause an additional dechanneling contribution by the accompanied Fourier frequency spectrum the channeled electron experiences.

Whether the discussed effects may decrease the dechanneling length by a factor in the order of two or three, or not, remains an open question. At the time being we somehow conservatively conclude that the prediction of the asymptotic dechanneling length on the basis of the Fokker-Planck equation represents an upper limit.

\acknowledgments


One of us (H.B.) acknowledges the hospitality of Professor Dr. Vadim Ivanov and Dr. Andrei Korol during his one month stay at the Peter The Great St. Petersburg Polytechnic University in April and June 2017. Many fruitful discussions, in particular with Andrei Korol, were of crucial importance for this work and are gratefully acknowledged.

This work has been supported by the European Commission (the PEARL Project within the H2020-MSCA-RISE-2015 call, GA 690991).



\bibliographystyle{JHEP}
\bibliography{bibfileRREPS17}

\providecommand{\href}[2]{#2}\begingroup\raggedright\begin{thebibliography}{10}

\bibitem{KorB16}
A.~V. Korol, V.~G. Bezchastnov, G.~B. Sushko and A.~V. Solov'yov,
  \emph{{Simulation of channeling and radiation of 855 MeV electrons and
  positrons in a small-amplitude short-period bent crystal}},
  \href{https://doi.org/http://dx.doi.org/10.1016/j.nimb.2016.09.021}{\emph{Nuclear
  Instruments and Methods in Physics Research B} {\bfseries 387} (2016) 41}.

\bibitem{KorB17}
A.~V. Korol, V.~G. Bezchastnov and A.~V. Solov'yov, \emph{{Channeling and
  radiation of the 855 MeV electrons enhanced by the re-channeling in a
  periodically bent diamond crystal}},
  \href{https://doi.org/10.1140/epjd/e2017-80113-y}{\emph{THE EUROPEAN PHYSICAL
  JOURNAL D} {\bfseries 71} (2017) 174}.

\bibitem{MazB14}
A.~Mazzolari, E.~Bagli, L.~Bandiera, V.~Guidi, H.~Backe, W.~Lauth et~al.,
  \emph{{Steering of a Sub-GeV Electron Beam through Planar Channeling Enhanced
  by Rechanneling}},
  \href{https://doi.org/10.1103/PhysRevLett.112.135503}{\emph{Physical Review
  Letters} {\bfseries 112} (2014) 135503}.

\bibitem{WisU16}
T.~N. Wistisen, U.~I. Uggerh{\o}j, U.~Wienands, T.~W. Markiewicz, R.~J. Noble,
  B.~C. Benson et~al., \emph{{Channeling, volume reflection, and volume capture
  study of electrons in a bent silicon crystal}},
  \href{https://doi.org/10.1103/PhysRevAccelBeams.19.071001}{\emph{Physical
  Review Accelerators and Beams} {\bfseries 19} (2016) 071001 11 pages}.

\bibitem{LauB08}
W.~Lauth, H.~Backe, P.~Kunz and A.~Rueda, \emph{{Channeling Experiments with
  Electrons at the Mainz Microtron MAMI}},  in \emph{Charged and Neutral
  Particles Channeling Phenomena, Channeling 2008, Proceedings of the 51st
  Workshop of the INFN ELOISATRON Project, Erice, Italy 15 October - 1 November
  2008}, S.~B. Dabagov, L.~Palumbo and A.~Zichichi, eds., {The Science and
  Culture Series - Physics}, (World Scientific Publishing Co. Pte. Ltd., 5 Toh
  Tuck Link, Singapore 596224), pp.~335--342, World Scientific: New Jersey,
  London, Singapore, Beijing, Shanghai, Hong Kong, Taipei, Chennai, 2008.

\bibitem{LauB10}
W.~Lauth, H.~Backe, P.~Kunz and A.~Rueda, \emph{{Channeling Experiments with
  Electrons at the Mainz Microtron MAMI}},
  \href{https://doi.org/10.1142/S0217751X10049980}{\emph{International Journal
  of Modern Physics A} {\bfseries 25, Supplement 1} (2010) 136}.

\bibitem{BacL08}
H.~Backe, P.~Kunz, W.~Lauth and A.~Rueda, \emph{{Planar channeling experiments
  with electrons at the 855 MeV Mainz Microtron MAMI}},
  \href{https://doi.org/http://dx.doi.org/10.1016/j.nimb.2008.05.012}{\emph{Nuclear
  Instruments and Methods in Physics Research B} {\bfseries 266} (2008) 3835 }.

\bibitem{BacK12}
H.~Backe, D.~Krambrich, W.~Lauth, K.~Andersen, J.~L. Hansen and U.~I.
  Uggerh{\o}j, \emph{Radiation emission at channeling of electrons in a
  strained layer {Si$_{1-x}$Ge$_x$} undulator crystal},
  \href{https://doi.org/http://dx.doi.org/10.1016/j.nimb.2013.03.047}{\emph{Nuclear
  Instruments and Methods in Physics Research B} {\bfseries 309} (2013) 37 }.

\bibitem{BacL15}
H.~Backe and W.~Lauth, \emph{{Channeling experiments with sub-GeV electrons in
  flat silicon single crystals}},
  \href{https://doi.org/http://dx.doi.org/10.1016/j.nimb.2015.03.077}{\emph{Nuclear
  Instruments and Methods in Physics Research B} {\bfseries 355} (2015) 24}.

\bibitem{GouS88}
M.~Gouanere, D.~Sillou, M.~Spighel, N.~Cue, M.~J. Gaillard, R.~G. Kirsch
  et~al., \emph{{Planar channeling radiation from 54-110-MeV electrons in
  diamond and silicon}}, {\emph{Physical Review B} {\bfseries 37} (1988) 4352
  }.

\bibitem{Aza13}
B.~Azadegan, \emph{{A Mathematica package for calculation of planar channeling
  radiation spectra of relativistic electrons channeled in a diamond-structure
  single crystal (quantum approach))}},
  \href{https://doi.org/10.1016/j.cpc.2012.11.013}{\emph{Computer Physics
  Communications} {\bfseries 184} (2013) 1064}.

\bibitem{DoyT67}
P.~A. Doyle and P.~S. Turner, \emph{{Relativistic Hartree-Foek X-ray and
  Electron Scattering Factors}}, {\emph{Acta Crystallographica A} {\bfseries
  24} (1968) 390}.

\bibitem{PanK87}
R.~Pantell, J.~Kephard, R.~Klein, H.~Park, B.~Berman and S.~Datz, \emph{{The
  Study of Material Properties Using Channeling Radiation}},  in
  \emph{Relativistic Channeling, Proceedings of a NATO Advanced Research
  Workshop on Relativistic Channeling, March 31 - April 4, 1986 at Villa Del
  Mare, Acquafredda di Maratea, Italy}, J.~R.A.~Carrigan and J.~Ellison, eds.,
  {NATO Advanced Science Institutes Series (ASI)}, (A Division of Plenum
  Publishing Corporation, 233 Spring Street, New York, N.Y. 10013),
  pp.~435--453, Plenum Press, New York and London, 1987.

\bibitem{PatS87}
A.~Pathak and S.~Satpathy, \emph{{Crystal Potentials from Channeling}},  in
  \emph{Relativistic Channeling, Proceedings of a NATO Advanced Research
  Workshop on Relativistic Channeling, March 31 - April 4, 1986 at Villa Del
  Mare, Acquafredda di Maratea, Italy}, J.~R.A.~Carrigan and J.~Ellison, eds.,
  {NATO Advanced Science Institutes Series (ASI)}, (A Division of Plenum
  Publishing Corporation, 233 Spring Street, New York, N.Y. 10013), pp.~455 --
  458, Plenum Press, New York and London, 1987.

\bibitem{BacK12A}
H.~Backe, D.~Krambrich, W.~Lauth, K.~K. Andersen, J.~L. Hansen and U.~I.
  Uggerh{\o}j, \emph{{Channeling and Radiation of Electrons in Silicon Single
  Crystals and Si$_{1-x}$Ge$_x$ Crystalline Undulators}},
  \href{https://doi.org/10.1088/1742-6596/438/1/012017}{\emph{Journal of
  Physics: Conference Series} {\bfseries 438} (2013) 012017}.

\bibitem{BaiK98}
V.~N. Baier, V.~M. Katkov and V.~M. Strakhovenko, \emph{{Electromagnetic
  Processes at High Energies in Oriented Single Crystals}}. World Scientific,
  Singapore, New Jersey, London, HongKong, {World Scientific Publishing Co.
  Pte. Ltd, P O Box 128, Farrer Road, Singapore 912805}, 1998.

\bibitem{BelT81}
V.~V. Beloshitsky and C.~G. Trikalinos, \emph{{Passage and radiation of
  relativistic channeled particles}}, {\emph{Radiation Effects} {\bfseries 56}
  (1981) 71}.

\bibitem{KumK89}
M.~A. Kumakhov and F.~F. Komarov, \emph{{Radiation from charged particles in
  solids}}. American Institute of Physics New York, {}, 1989.

\bibitem{Pat17}
C.~Patrignani and et~al. (Particle Data~Group), \emph{{Table 6. Atomic and
  Nuclear Properties of Materials}},
  \href{https://doi.org/http://pdg.lbl.gov/2017/reviews/rpp2017-rev-atomic-nuclear-prop.pdf}{\emph{Chin.
  Phys. C} {\bfseries 40} (2016 and 2017 update) 100001}.

\bibitem{SeaS91}
V.~F. Sears and S.~A. Shelley, \emph{{Debye-Waller Factor for Elemental
  Crystals}}, {\emph{{Acta Crystallographica A}} {\bfseries 47} (1991) 441}.

\bibitem{Ioffe}
{Vadim Siklitsky and Alexei Tolmatchev (Ioffe Physico-Technical Institute)},
  ``{Electronic archive: New Semiconductor Materials. Characteristics and
  Properties}.'' \url{http://www.ioffe.ru/SVA/NSM/Semicond/Diamond/basic.html},
  1998-2001.

\bibitem{Matprop}
``{New Semiconductor Materials. Biology systems. Characteristics and
  Properties}.'' \url{http://www.matprop.ru/Diamond_basic}.

\bibitem{RosG41}
B.~Rossi and K.~Greisen, \emph{{Cosmic-Ray Theory}}, {\emph{{Review of Modern
  Physics}} {\bfseries 13} (1941) 240}.

\bibitem{LynD91}
G.~Lynch and O.~Dahl, \emph{{Approximations to multiple Coulomb scattering}},
  {\emph{Nuclear Instruments and Methods in Physics Research B} {\bfseries 58}
  (1991) 6}.

\bibitem{NetG94}
U.~Nething, M.~Galemann, H.~Genz, M.~Hofer, P.~Holfmann-Stascheck, J.~Hormes
  et~al., \emph{{Intensity of Electron Channeling Radiation and Occupation
  Lengths in Diamond Crystals}}, {\emph{Physical Review Letters} {\bfseries 72}
  (1994) 2411}.

\bibitem{KepP89}
J.~O. Kephart, R.~H. Pantell, B.~L. Berman, S.~Datz, H.~Park and R.~K. Klein,
  \emph{{Measurement of the occupation lengths of channeled 17-MeV electrons
  and 54-MeV electrons and positrons in silicon by means of channeling
  radiation}}, \href{https://doi.org/10.1103/PhysRevB.40.4249}{\emph{Physical
  Review B} {\bfseries 40} (1989) 4249 }.

\end{thebibliography}\endgroup

\end{document}